\documentclass[fianl,1p,times]{elsart}

\usepackage{graphicx}
\usepackage{amssymb}
\journal{Physica A}

\begin{document}
\begin{frontmatter}


\title{An Evolving model of online bipartite networks}

\author[inst1,inst2]{Chu-Xu Zhang,}
\author[inst1,inst2,inst3]{Zi-Ke Zhang,}
\author[inst1]{Chuang Liu}
\corauth{Corresponding Author: zhangzike@gmail.com}

  \address[inst1]{ Institute of Information Economy, Hangzhou Normal University, Hangzhou 310036, PRC}
  \address[inst2]{ Web Sciences Center, University of Electronic Science and Technology of China, Chengdu 610054, PRC}
  \address[inst3]{ Department of Physics, University of Fribourg, Chemin du Mus\'ee 3, 1700 Fribourg,Switzerland}



\begin{abstract}
Understanding the structure and evolution of online bipartite networks is a significant task since they play a crucial role in various e-commerce services nowadays. Recently, various attempts have been tried to propose different models, resulting in either power-law or exponential degree distributions.
However, many empirical results show that the user degree distribution actually follows a shifted power-law distribution, so-called \emph{Mandelbrot law}, which cannot be fully described by previous models. In this paper, we propose an evolving model, considering two different user behaviors: random and preferential attachment. Extensive empirical results on two real bipartite networks, \emph{Delicious} and \emph{CiteULike}, show that the theoretical model can well characterize the structure of real networks for both user and object degree distributions. In addition, we introduce a structural parameter $p$, to demonstrate that the hybrid user behavior leads to the shifted power-law degree distribution, and the region of power-law
tail will increase with the increment of $p$. The proposed model might shed some lights in understanding the underlying laws governing the structure of real online bipartite networks.

\end{abstract}


\end{frontmatter}

\section{Introduction}
The past decade has witnessed a great explosion of studying and understanding the underlying mechanisms of various real-life networks, ranging from the Internet, scientific collaboration networks, protein networks to social networks, etc \cite{Albert200201, Dorogovtsev200201, Newman200301, Boccaletti200601, Costa200701, Arenas200801, Castellano200901}. Although they respectively have their own properties and characteristics, empirical analyses show that many common characteristics and phenomena can be discovered from networks with such a wide-range functions, e.g. a small average distance between nodes, a large clustering coefficient \cite{Watts199801}, power-law degree distribution \cite{Barabasi199901} and community structures \cite{GirvanM200201} of the emerging structure.
Recently, studies on the mathematics of networks have been driven largely by those observed empirical properties of real networks, as well as network dynamics. However, many pioneering works in this area focus on designing evolutionary models of unipartite neworks which only have one kind of nodes, such as Erd{\H{o}}s-R{\'e}nyi network \cite{ErdosP1960}, Watt-Strogatz network \cite{Watts199801}, Barab{\'a}si-Albert network \cite{Barabasi199901}, as well as many extensive variants considering different factors (e.g. aging effect \cite{Dorogovtsev2000, Dorogovtsev2007} and social impact \cite{Jin2001, Newman2003, Castellano2009}).

Recently, with the advent of Web 2.0 and affiliated applications, the family of \emph{Networks} also has received many new members. One example is the bipartite network which involves two different kinds of nodes with different functions \cite{Peltomaki2006, Goldstein2005, Shang201002}. Different from traditional networks, the nodes in a pure bipartite network can be divided into two independent communities, where edges are only allowed to exist between different communities. Nowadays, this bipartite network is widely applied in both online platforms (e.g. online services where users view/purchase products \cite{WangDH2006, Peruani2007, Hidalgo2009}, or listen to music \cite{Lambiotte2005}), biology \cite{Ergun2002, Aittokallio2006, Maayan2007,Goh2007} and medical science \cite{Yildirim2007, Nacher2008, Barabasi2011} and theoretical studies \cite{Strogatz2001,Watts2004,Guillaume2004, Mucha2010}. There is also a vast class of researches that have recently reported many universal properties in unipartite networks, such as power-law degree distribution and correlation \cite{Peltomaki2006, Shang201002} and community structure \cite{Mucha2010, Lind200502, Guimera2007, Barber2007,ZhangP2008}, could also be found in bipartite networks. Consequently, it has attracted an increasing attention from scientific community due to its wide application and bright prospect in characterizing the essential properties of real networks. The first and natural attempt is to project the bipartite network to a corresponding unipartite network and using methods for traditional networks \cite{Lambiotte200502,Kossinets2006,Lehmann2008, Mukherjee2011}.

However, it is argued that such one-mode projection ignores much informative structure and relationship, subsequently, it would give unreliable or incorrect results \cite{Newman2002, Ramasco2004, Guimera2007}. Therefore, a more common approach is to keep the original bipartite structure, investigate both its specific and common properties, and try to uncover the underlying mechanism driving the emergence of this two-mode network. Newman \emph{et al.} used the random graph model to describe social networks of both unipartite and bipartite relations \cite{Newman2002}. Using generating functions \cite{Newman200101}, they concluded that the clustering and average degree of real affiliation networks, as one typical kind of bipartite networks, agreed well with the theoretical prediction. Lambiotte \emph{et al.} proposed a personal identification and community imitation (PICI) based model to consider both effects of collective behavior and personalization \cite{Lambiotte2005}. This model generated an exponential and power-law degree distribution for music groups and owners, respectively. Sood and Redner introduced the voter model on networks of power-law degree distributions with and without degree correlation, both of which showed the consensus time was greatly dependent on the value of exponent \cite{Sood2005}. Noh \emph{et al.} demonstrated that different mechanisms would generate different shape of degree distributions in group selection systems \cite{Noh2005}. That is to say, a random selection process would result in an exponential distribution of the activity degree, otherwise a power-law distribution of group size and activity degree would arise from the resultant force of preferential selection and fixed-probability creation.  Sneppen \emph{et al.} proposed a minimalistic model of directed bipartite network, and a self-organization phenomenon was observed by a dynamical reconnection process \cite{Sneppen2007}. Similar result was also found in collaboration bipartite networks via preferential attachment of actors' degree \cite{Ramasco2004}. Hence, this model only reproduced that one kind of node followed power-law but neglecting outputs of the other side of nodes. Saavedra \emph{et al.} introduced two mechanisms, specialization and interaction, would produce exponential degree distribution for both sides \cite{Saavedra2008}. In addition, they found this bipartite cooperation can well characterize the structure of both ecological and organization networks.

In this paper, we focus on studying the degree distribution of online bipartite networks where users view/choose/select objects (e.g. bookmarks, music, movies), as well as the underlying mechanisms. Despite many previous studies demonstrated that both exponential and power-law degree distribution could be obtained by corresponding models, empirical analysis of some online bipartite networks shows that the user degree distribution actually follows shifted power-law, so-called \emph{Mandelbrot law} \cite{Mandelbrot1965, Ren2012}, instead of purely exponential or power-law decay, while the object degree distribution always obeys power-law \cite{Shang201002, ZhangZK201002}, and it can not be fully explained by previous models. Therefore, We propose an evolutionary model to consider the proactive selection activity of users and the passive pattern of objects. Theoretical analysis shows that the present model can not only well reproduce the two different degree distribution, but also find good agreements of two real-world data sets, \emph{Delicious}\footnote{http://www.delicious.com/} and \emph{CiteULike}\footnote{http://www.citeulike.com}. In addition, we find that the structural parameter $p$, determines the transformation from exponential to power-law decay of the user degree distribution.
\section{Model}
In this section, we shall propose an evolving model to uncover the growing dynamics of online bipartite networks. Here, we mainly consider two mechanisms: random and preferential attachment. In particular, we assume there are two kind of online behaviors for users: she can either randomly choose an object or pick up an item according to its popularity. On one hand, considering a new user involving in the system, it would be difficult for her to select a suitable object from numerous candidates. One reasonable action she would take is to choose a popular item since other users also like it. On the other hand, old users who have devoted much time in playing the online platform, would know to find their own favorites and thus are likely to select personalized (hence might be less popular) items. That is to say, users are very proactive in performing online activities. In \cite{LiuZ2002, ZhangZZ2007}, they reported such a hybrid behavior would result in an intermediate status between power-law and exponential distribution. By contrast, objects in online systems are always in a passive pattern, hence do not have any choice but waiting to be selected to gain popularity. Therefore, we assume objects always grow based on preferential attachment in our model.

We begin our study with some related definitions of bipartite graph that we will analyze. The bipartite graph can be represented by
$G=(U,O,E)$, where $U$ and $O$ are two disjoint sets of nodes, respectively representing users and objects, and $E\subseteq U\times O$ is the set of edges. The difference with classical graph lies in the fact that edges exist only between user vertices and object vertices. The model starts from an initial bipartite network: there exist $u_0$ nodes in $U$, $o_0$ nodes in $O$ and $e_0$ edges in set $E$. Given a user $i$ in $U$ and an object $j$ in $O$, denote $k_i$ as the degree of $i$ and $l_j$ as the degree of $j$ in the bipartite network. Then, $e_0=\sum\limits_i k_i=\sum\limits_j l_j(k_i,l_j\geq1, i=1,2,...,u_0, j=1,2,...,o_0)$. There are totally $N=u_0+t$ users and $M=o_0+t$ objects in the model at time $t$. Consequently, the model can be described as following:

\begin{itemize}
\item adding a new user:  Connect the new user node to $m$ different nodes already in $O$ by preferential probability $\frac{l_j}{\sum\limits_{j=1}^{o_0+t-1} l_j}$.
\item adding a new object:  Link the new node to $n$ different nodes already in $U$ by preferential probability $\frac{k_i}{\sum\limits_{i=1}^{u_0+t-1} k_i}$.
\item edges evolving randomly: Two kinds of old nodes are connected by $c$ edges, which are chosen as: users are selected randomly with  probability $\frac{1}{u_0+t}$, while objects in $O$ are selected by preferential probability $\frac{l_j}{\sum\limits_{j=1}^{o_0+t-1} l_j}$.
\item edges evolving by preferential attachment:  Two kinds of old nodes are connected by $b$ edges, which are chosen as: users are selected by preferential probability $\frac{k_i}{\sum\limits_{i=1}^{u_0+t-1} k_i}$, and objects are also selected by preferential probability $\frac{l_j}{\sum\limits_{j=1}^{o_0+t-1} l_j}$.
\end{itemize}

\begin{figure}[htb]
  \centering
  \includegraphics[width=14.5cm,height=6cm]{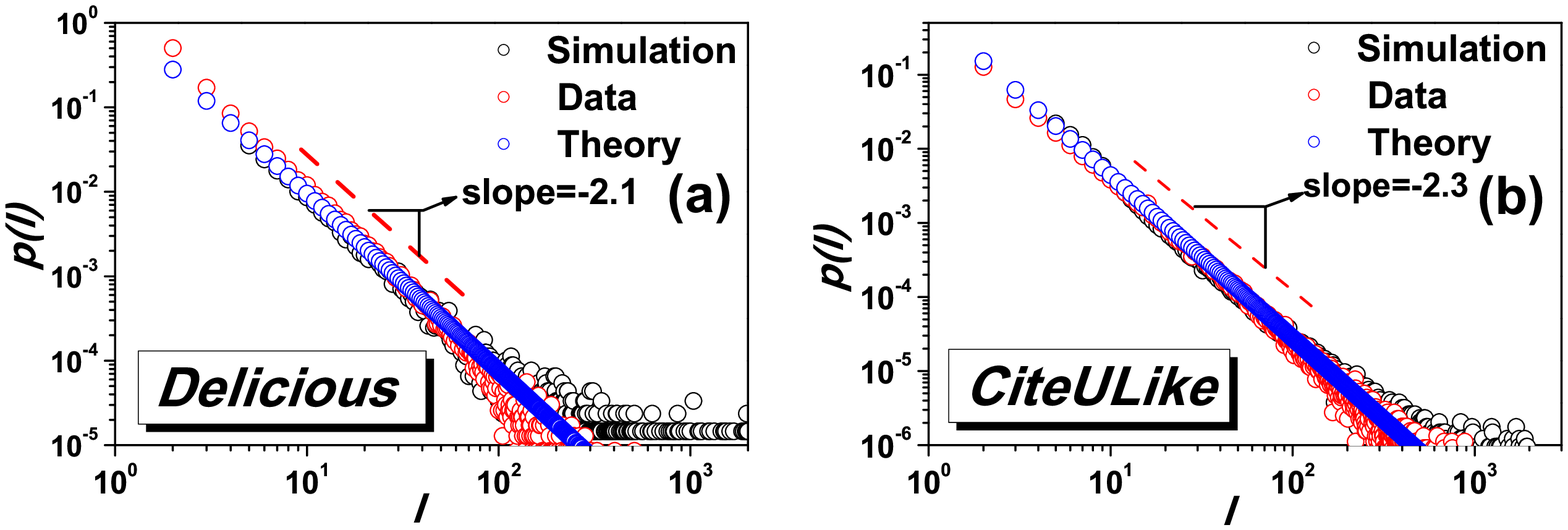}
  \caption{(Color online)\label{Fig:fig2} Object degree distribution in a log-log scale of $Delicious$ (left) and $CiteULike$ (right). The parameters used for theory and simulation are set as: 1.) $n=m=5$, $b=65$ and $c=30$ for Delicious; 2.) $n=m=c=5$ and $b=55$ for CiteULike. }
\end{figure}

\section{Analytical Analysis}

\subsection{Object degree distribution}
From the aforementioned model description, we can write the dynamics of degree for object $Oj$
\begin{equation}
\centering
\label{EQ:eq1}
\frac{\partial l_j}{\partial t}=m\frac{l_j}{\sum\limits_{j=1}^{o_0+t-1} l_j}+c\frac{l_j}{\sum\limits_{j=1}^{o_0+t-1} l_j}+b\frac{l_j}{\sum\limits_{j=1}^{o_0+t-1} l_j},
\end{equation}
where $\sum\limits_{j=1}^{o_0+t-1}l_j=\langle l\rangle M$, $M=o_0+t$, $\langle l\rangle=\frac{(m+n+c+b)t+e_0}{o_0+t}$. Then Eq. \ref{EQ:eq1} is approximated to
\begin{equation}
\label{EQ:eq2}
\frac{\partial l_j}{\partial t}=\frac{wl_j}{vt},
\end{equation}
where $w$=$m+c+b$, $v$=$m+n+c+b$, t$\gg$m,n,c,b and $i=1,2,...,t$.

The initial degree of node $j$ satisfies $l_j(t_j)$=$n$, where $t_j$ represents the time that node $j$ is added into O. Therefore we obtain following equation by solving Eq. \ref{EQ:eq2} \cite{Albert200201}
\begin{equation}
\label{EQ:eq3}
l_j(t)=n(\frac{t}{t_i})^{\frac{w}{v}}
\end{equation}
Let $l_j(t)<l$,then $t_i>t(\frac{n}{l})^{\frac{v}{w}}$. So the cumulative probability $P(l_j(t)<l)$ can be denoted by $P(t_i>t(\frac{n}{l})^{\frac{v}{w}})$, such that
\begin{equation}
\label{EQ:eq4}
P(l_j(t)<l)=P(t_i>t(\frac{n}{l})^{\frac{v}{w}}).
\end{equation}
In the model, all nodes are added into network with the same time interval, which means
\begin{equation}
\label{EQ:eq5}
p(t_j)=\frac{1}{o_0+t}.
\end{equation}
Integrating Eq. \ref{EQ:eq4} and Eq. \ref{EQ:eq5}, we can obtain the cumulative probability
\begin{equation}
\label{EQ:eq6}
p(l_j(t)<l)=p(t_j>{t\frac{n}{l}^{\frac{v}{w}}})=1-\frac{t}{o_0+t}(\frac{l}{n})^{-\frac{v}{w}}.
\end{equation}
Finally, with assuming as $t\gg m,n,c,b$, the object degree distribution can be written
\begin{equation}
\label{EQ:eq7}
p(l)=\frac{\partial p(l_j(t)<l)}{\partial l}\approx \frac{v}{w}n^{\frac{v}{w}}l^{-\frac{v}{w}-1}.
\end{equation}
From Eq. \ref{EQ:eq7}, it is can be found that the object degree distribution accords with power-law distribution, with exponent $\gamma_l=1+\frac{v}{w}$.
\subsection{User degree distribution}
Similar to the theoretical analysis of object degree distribution, the dynamics of user $u_i$ can be written as
\begin{equation}
\label{EQ:eq8}
\frac{\partial k_i}{\partial t}=n\frac{k_i}{\sum\limits_{i=1}^{u_0+t-1} l_j}+c\frac{1}{N}+b\frac{k_i}{\sum\limits_{k=1}^{u_0+t-1} l_j},
\end{equation}
where $\sum\limits_{i=1}^{u_0+t-1}k_i$=$\langle k \rangle N$, $N=u_0+t$, $\langle k\rangle=\frac{(m+n+c+b)t+e_0}{u_0+t}$. Then Eq. \ref{EQ:eq8} is approximated to
\begin{equation}
\label{EQ:eq9}
\frac{\partial k_i}{\partial t}=\frac{uk_i}{vt}+\frac{c}{N},
\end{equation}
where $u$=$n+b$, $v$=$m+n+c+b$, t$\gg$m,n,c,b and $i=1,2,...,t$.

Since the initial degree of all users satisfies $k_i(t_i)=m$, where $t_i$ represents the time user $u_i$ is added into $U$. Then we get following equation by solving Eq. \ref{EQ:eq9}
\begin{equation}
\label{EQ:eq10}
k_i(t)=\frac{(\frac{t}{t_i})^{\frac{u}{v}}(cv+mu)-cv}{u}.
\end{equation}
Substitute $p(t_i)=\frac{1}{u_0+t}$ into Eq. \ref{EQ:eq10}, we will get the cumulative probability
\begin{equation}
\label{EQ:eq11}
p(k_i(t)<k)=1-\frac{t}{u_0+t}(\frac{cv+ku}{cv+mu})^{-\frac{v}{u}}.
\end{equation}
So the user degree distribution function is finally achieved by assuming $t\gg m,n,c,b$
\begin{equation}
\label{EQ:eq12}
p(k)=\frac{\partial p(k_i(t)<k)}{\partial k}\approx(cv+mu)^{\frac{v}{u}}v(cv+ku)^{-\frac{v}{u}-1}.
\end{equation}
From Eq. \ref{EQ:eq12}, we know that the user degree distribution is a shifted power-law distribution \cite{LiuZ2002, ZhangZZ2007}, which is also familiar as Mandelbort law \cite{Mandelbrot1965, Ren2012}.
\section{Results \& Analysis}
In this section, we use two data sets to evaluate the proposed model. The first one is \emph{Delicious}, one of the most popular social bookmarking web sites, which allows users not only to store and organize personal bookmarks, but also to look into users' collection and find what they might be interested in \cite{ZhangZK201001}. The other is from $CiteULike$, which also has similar characterizations with $Delicious$. The objects are common website and publication URL for Delicious and CiteULike, respectively. Table. \ref{Lab:lab1} shows the basic statistical properties of the two data sets.

\subsection{Degree distributions}

Fig. \ref{Fig:fig2} reports the object degree distribution result. It can be seen that the simulation and analytical results fit will with the real data. In addition all the object-degree distributions are power-law, as $p(l)\propto l^{-\gamma_l}$, with $\gamma_l=$ 2.1 and 2.3 for $Delicious$ and $CiteULike$, respectively.

\begin{figure}[htb]
  \centering
  \includegraphics[width=14.5cm,height=6cm]{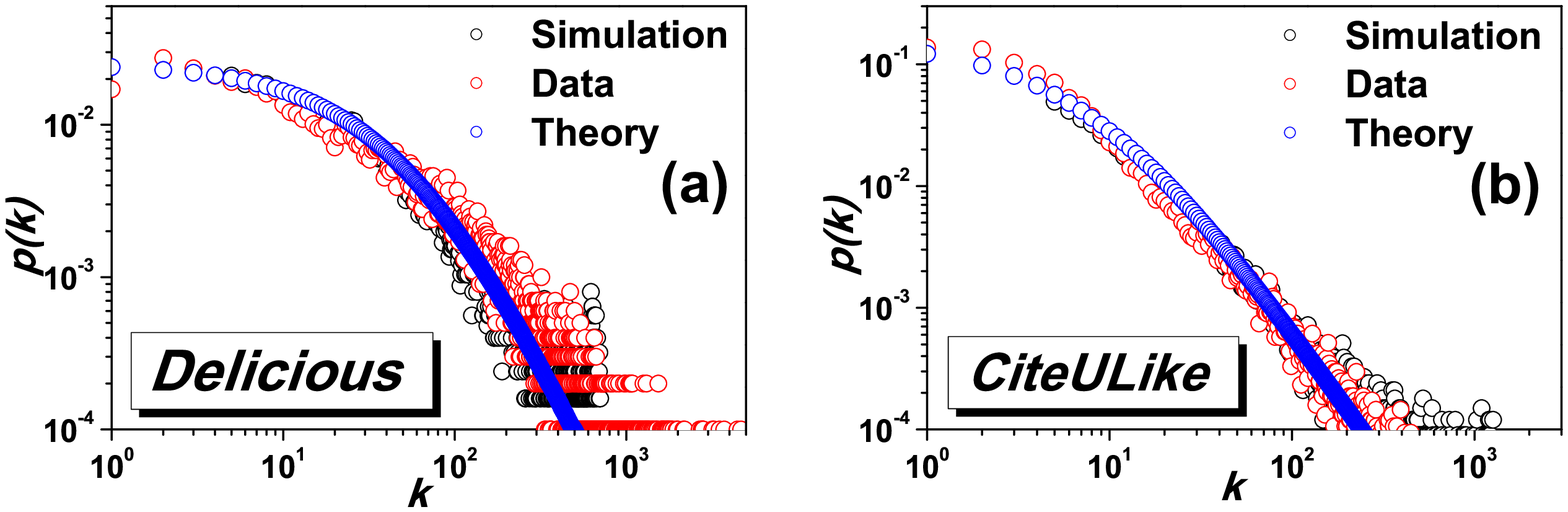}
  \caption{(Color online)\label{Fig:fig3} User degree distribution in a log-log scale of $Delicious$ (left) and $CiteULike$ (right). The corresponding parameters are the same as Fig. \ref{Fig:fig2}. }
\end{figure}

Fig. \ref{Fig:fig3} illustrates the user degree distribution. Again, we find good agreements among the simulation, analytical and empirical results. Therefore, the present model can qualitatively accurate for modeling the general real-world networks by assuming users' mixture behavior. The degree distributions for all users are similar to shifted power-law distribution $p(k)\propto (k_0k+c_0)^{-\gamma_u}$, with $\gamma_u=$ 2.5 and 2.2 for $Delicious$ and $CiteULike$, respectively. $k_0$ and $c_0$ are constants.

\begin{table}[htbp]
\centering \caption{\label{Lab:lab1} Basic statistical properties of the $Delicious$ and $Citeulike$. $|U|$, $|O|$ and $|E|$ denote the number of users, objects and edges, respectively. $\rho=\frac{|E|}{|U|\times|O|}$ denotes the sparsity of the data.}
\begin{tabular}{cccccc}  \hline \hline Data set & $|U|$ & $|O|$ & $|E|$ & $\rho$
\\ \hline
\emph{$Delicious$} & 9,998 & 232,657 & 123,995 &   5.305 $\times 10^{-4}$ \\
\emph{$Citeulike$} &42,801&397,536  &7,083,253 &  4.163 $\times 10^{-4}$\\
\hline \hline
\end{tabular}
\end{table}

\subsection{Understanding the effects of random and preferential attachment}
\begin{figure}[htb]
  \centering
  \includegraphics[width=12.5cm,height=16cm]{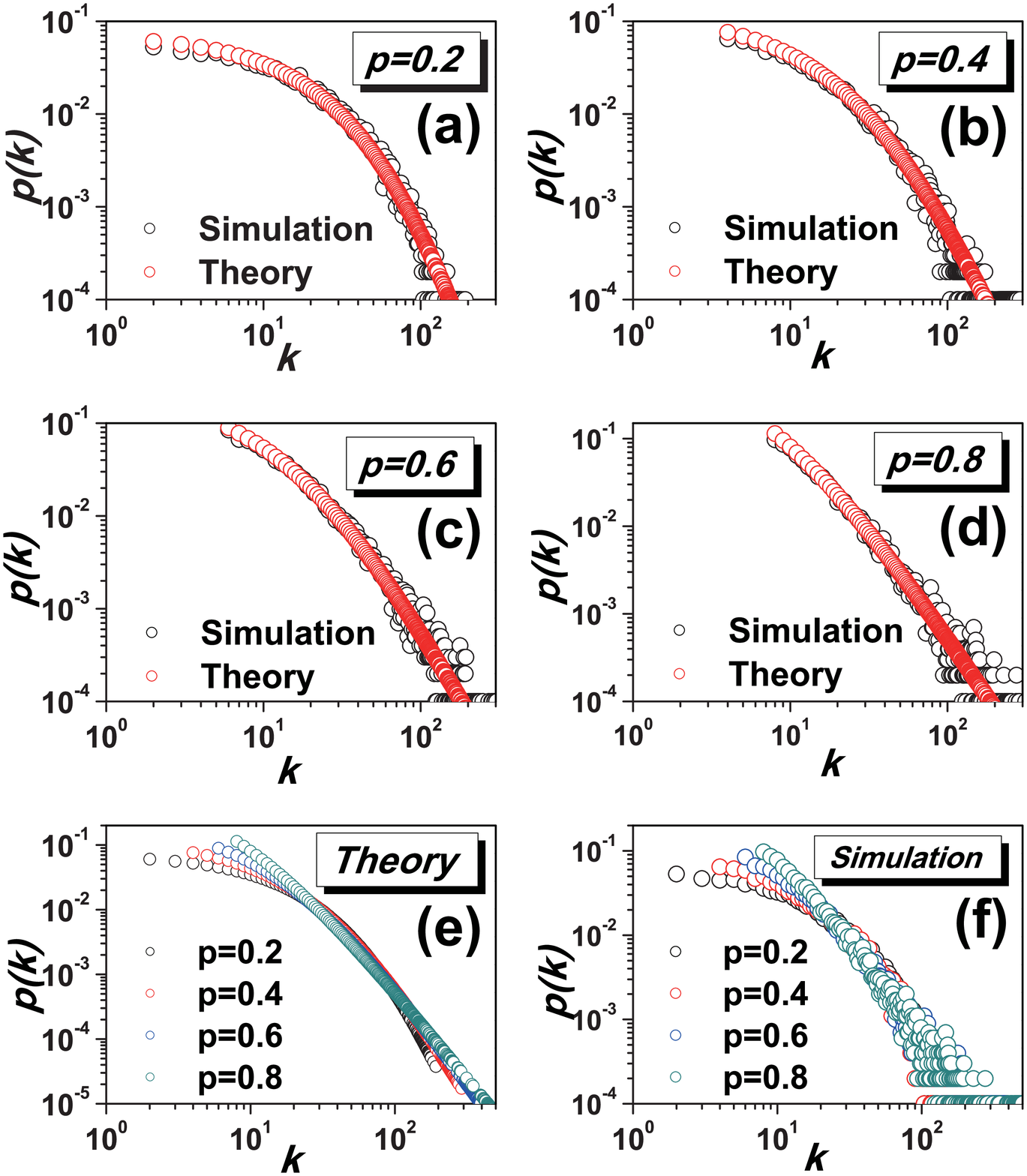}
  \caption{(Color online)\label{Fig:fig4} The theoretical and simulation user degree distributions in log-log scale for different $p$, including (a) $p$=0.2; (b) $p$=0.4; (c) $p$=0.6; (d) $p$=0.8. In addition, (e) and (f) compare the theoretical and simulation results for different $p$, respectively. The parameters are set as: 1.) $n=m=b=2$ and $c=16$ for $p=0.2$; 2.) $n=m=b=4$ and $c=12$ for $p=0.4$; 3.) $n=m=b=6$ and $c=8$ for $p=0.6$; 4.) $n=m=b=8$ and $c=4$ for $p=0.8$. }
\end{figure}

From the analysis of network estimation, the user degree distribution is determined together by both preferential and random linking mechanisms. In order to further understand the effects of these two mechanisms, we introduce a structural parameter, $p \in [0,1]$, to quantify different weights of them. Denote $p$ as the weight of preferential mechanism, and $1-p$ refers to random choosing mechanism. According to the model description, we have $p=\frac{n+b}{n+b+c}$. Fig. \ref{Fig:fig4} shows both theoretical and simulation results of the user degree distribution for different $p$.

As shown in Fig. \ref{Fig:fig4}(e) and \ref{Fig:fig4}(f), an obvious correlation between $p$ and the user degree distribution is observed. In addition, the scale-free region increases with the increment of $p$, which indicates that $p$ indeed can characterize the different structures driven by the two mechanisms. In particular, for the extreme cases:
\begin{itemize}
  \item $p=1$. It means $c=0$. Thus, Eq. \ref{EQ:eq12} will degenerate to $p(k)\propto k^ {-(2+\frac{m}{n+b})}$, which is a pure scale-free degree distribution.
  \item $p=0$. It means $n=b=0$ or $c\rightarrow \infty$. In this case, Eq. \ref{EQ:eq8} will degenerate to
   \begin{equation}
       \label{EQ:eq13}
       \frac{\partial k_i}{\partial t}=c\frac{1}{N}.
   \end{equation}
    Apparently, Eq. \ref{EQ:eq13} shows that users will randomly choose objects. We can easily obtain its solution
       \begin{equation}
       \label{EQ:eq14}
       p(k)\approx\frac{1}{c}e^{\frac{-k+m}{c}} \propto e^{-\frac{k}{c}}.
   \end{equation}
    Therefore, Eq. \ref{EQ:eq14} suggests that the user degree distribution will follow an exponential form in the extreme case $p=0$.
\end{itemize}

Otherwise, a shifted power-law decay will be observed for $p \in (0,1) $. The form of shifted power-law distribution, now well known as Mandelbrot law, is $p(k)\propto  (k_0k+c_0)^{-\gamma_u}$, where $k_0$ and $c_0$ are constants, and $\gamma_u$ is the characteristic exponent.


\subsection{Conclusions and Discussion}

Previous models about evolving bipartite networks usually lead to power-law degree distribution for both of users and objects, which conflicts with the properties of some real networks, of which user degree distribution is shifted power-law distribution, so-called Mandelbrot law. In this paper, we propose an evolving model, trying to characterize the hybrid user behaviors. The proposed model considers that users' actions are determined by both random and preferential mechanisms, and objects are selected mainly by preferential mechanism. Results of real data, theory and simulation are well fitted with each other. In addition, we also compare the weights of the two different mechanisms, and find out that a clear correlation between the structural parameter and the shape of user degree distribution. Our proposed model might shed some lights in understanding the underlying laws governing the structure of real online bipartite networks.

\section{Acknowledgements}
This work was partially supported by the National Natural Science Foundation
of China (Grant Nos. 11105024, 11105040, 61103109), and the Zhejiang Provincial Natural
Science Foundation of China under Grant No. LY12A05003. ZKZ acknowledges the start-up foundation of Hangzhou Normal University and the Pandeng Project.


\begin{thebibliography}{99}

\bibitem{Albert200201} R. Albert, A.-L. Barab\'{a}si, Rev. Mod. Phys., \textbf{74} (2002) 47.

\bibitem{Dorogovtsev200201}S. N. Dorogovtsev, J. F. F. Mendes, Adv. Phys., \textbf{51} (2002) 1079.

\bibitem{Newman200301}M. E. J. Newman, SIAM Rev., \textbf{45} (2003) 167.

\bibitem{Boccaletti200601}S. Boccaletti, V. Latora, Y. Moreno, M. Chavez, D.-U. Huang, Phys. Rep., \textbf{424} (2006) 175.

\bibitem{Costa200701}L. da F. Costa, F. A. Rodrigues, G. Traviesor, P. R. U. Boas, Adv. Phys., \textbf{56} (2007) 167.

\bibitem{Arenas200801}A. Arenas, A. D{\'\i}az-Guilera, J. Kurths, Y. Moreno, C. Zhou, Phys. Rep., \textbf{469} (2008) 93.

\bibitem{Castellano200901}C. Castellano, S. Fortunato, V. Loreto, Rev. Mod. Phys., \textbf{81} (2009) 591.

\bibitem{Watts199801}D. J. Watts, S. Strogatz, Nature, \textbf{393} (1998) 440.

\bibitem{Barabasi199901}A.-L. Barab{\'a}si, R. Albert, Science, \textbf{286} (1999) 509.

\bibitem{GirvanM200201}M. Girvan, M. E. J. Newman, Proc. Natl. Acad. Sci. U.S.A., \textbf{99} (2002) 7821.

\bibitem{ErdosP1960}P. R{\'e}nyi. A. Erd{\H{o}}s, Akad. Kiad{\'o}, 1960.

\bibitem{Dorogovtsev2000}S. N. Dorogovtsev, J. F. F. Mendes, Phys. Rev. E, \textbf{62} (2000) 1842.

\bibitem{Dorogovtsev2007}S. N. Dorogovtsev, J. F. F. Mendes, Europhys Lett., \textbf{52} (2007) 33.

\bibitem{Jin2001}S. N. Dorogovtsev, J. F. F. Mendes, Phys. Rev. E, \textbf{64} (2001) 046132.

\bibitem{Newman2003} M. E. J. Newman, J. Park, Phys. Rev. E, \textbf{68} (2003) 036122.

\bibitem{Castellano2009}C. Castellano, S. Fortunato, V. Loreto, Rev. Mod. Phys., \textbf{81} (2009) 591.

\bibitem{Peltomaki2006}M. Peltom{\"a}ki, M. Alava, J. Stat. Mech., (2006) P01010.

\bibitem{Goldstein2005}M. L. Goldstein, S. A. Morris, G. G. Yen, Phys. Rev. E, \textbf{71} (2005) 026108.

\bibitem{Shang201002}M.-S. Shang, L. L\"{u}, Y.-C. Zhang, T. Zhou, Europhys Lett., \textbf{90} (2010) 48006.

\bibitem{WangDH2006}D. Wang, L. Zhou, Z. D. Di, Physica A, \textbf{363} (2006) 359.

\bibitem{Peruani2007}F. Peruani, M. Choudhury, A. Mukherjee, N. Ganguly, Europhys Lett., \textbf{79} (2007) 28001.

\bibitem{Hidalgo2009}C. A. Hidalgo, R. Hausmann, Proc. Natl. Acad. Sci. U.S.A., \textbf{106} (2009) 10570.

\bibitem{Lambiotte2005}R. Lambiotte, M. Ausloos, Phys. Rev. E, \textbf{72} (2005) 066107.

\bibitem{Ergun2002}G. Erg{\"u}n, Physica A, \textbf{308} (2002) 483.

\bibitem{Aittokallio2006}T. Aittokallio, B. Schwikowski, Brief. Bioinform., \textbf{7} (2006) 243.

\bibitem{Maayan2007}A. Ma'ayan, S. L. Jenkins, J. Goldfarb, R. Iyengar, Mt.Sinai J. Med., \textbf{74} (2007) 27.

\bibitem{Goh2007}K. I. Goh, M. E. Cusick, D. Valle, B. Childs, M. Vidal, A.-L. Barab{\'a}si, Proc. Natl. Acad. Sci. U.S.A., \textbf{104} (2007) 8685.

\bibitem{Yildirim2007}M. A. Yildirim, K. I. Goh, M. E. Cusick, A.-L. Barab{\'a}si, Vidal. M, Nat. Biotechnol., \textbf{25} (2007) 1119.

\bibitem{Nacher2008}J. C. Nacher, J. M. Schwartz, BMC Pharmacol., \textbf{8} (2008) 5.

\bibitem{Barabasi2011}A. L. Barab{\'a}si, N. Gulbahce, J. Loscalzo, Nat. Rev. Genet., \textbf{12} (2011) 56.

\bibitem{Strogatz2001}S. H. Strogatz, Nature, \textbf{410} (2001) 268.

\bibitem{Watts2004} D. J. Watts, Ann. Rev. Sociol.,(2004) 243.

\bibitem{Guillaume2004}J. L. Guillaume, M. Latapy, Infor. Process. Lett., \textbf{90} (2004) 215.

\bibitem{Mucha2010} P. J. Mucha, T. Richardson, K. Macon, M. A. Porter, J. P. Onnela, Science, \textbf{328} (2010) 876.

\bibitem{Lind200502}P. G. Lind, M. C. Gonzalez, H. J. Herrmann, Phys. Rev. E, \textbf{72} (2005) 056127.

\bibitem{Guimera2007}R. Guimer{\`a}, M. Sales-Pardo, L. A. N. Amaral, Phys. Rev. E, \textbf{76} (2007) 036102.

\bibitem{Barber2007} M. J. Barber, Phys. Rev. E, \textbf{76} (2007) 066102.

\bibitem{ZhangP2008}P. Zhang, J. Wang, X. Li, M. Li, Z. Di, Y. Fan, Physica A, \textbf{387} (2008) 6869.

\bibitem{Lambiotte200502}R. Lambiotte, M. Ausloos, Phys. Rev. E, \textbf{72} (2005) 066117.

\bibitem{Kossinets2006}G. Kossinets, Social Netw., \textbf{28} (2006) 247.

\bibitem{Lehmann2008}S. Lehmann, M. Schwartz, L. K. Hansen, Phys. Rev. E, \textbf{78} (2008) 016108.

\bibitem{Mukherjee2011}A. Mukherjee, M. Choudhury, N. Ganguly, Physica A, \textbf{390} (2011) 3602.

\bibitem{Newman2002}M. E. J. Newman, D. J. Watts, S. H. Strogatz, Proc. Natl. Acad. Sci. U. S. A., \textbf{99} (2002) 2566.

\bibitem{Ramasco2004}J. J. Ramasco, S. N. Dorogovtsev, R. Pastor-Satorras, Phys. Rev. E, \textbf{70} (2004) 036106.

\bibitem{Newman200101}M. E. J. Newman, S. H. Strogatz, D. J. Watts, Proc. Natl. Acad. Sci. U. S. A., \textbf{98} (2001) 404.

\bibitem{Sood2005}V. Sood, S. Redner, Phys. Rev. Lett., \textbf{94} (2005) 178701.

\bibitem{Noh2005}J. D. Noh, H. C. Jeong, Y. Y. Ahn, H. Jeong, Phys. Rev. E, \textbf{71} (2005) 036131.

\bibitem{Sneppen2007}K. Sneppen, M. Rosvall, A. Trusina, P. Minnhagen, Europhys Lett., \textbf{67} (2007) 349.

\bibitem{Saavedra2008}S. Saavedra, F. Reed-Tsochas, B. Uzzi, Nature, \textbf{457} (2008) 463.

\bibitem{ZhangZK201002}Z.-K. Zhang, C. Liu, J. Stat. Mech., (2010) P10005.

\bibitem{Mandelbrot1965} B. Mandelbrot, New York: Baisc Books Publishing Co. (1965).

\bibitem{Ren2012} X.-Z. Ren, Z.-M. Yang, B.-H. Wang, T. Zhou, Chin. Phys. Lett., \textbf{29} (2012) 038904.

\bibitem{LiuZ2002}Z. Liu, Y. C. Lai, N. Ye, P. Dasgupta, Phys. Lett. A, \textbf{303} (2002) 337.

\bibitem{ZhangZZ2007}Z. Zhang, L. Rong, B. Wang, S. Zhou, J. Guan, Physica A, \textbf{380} (2007) 639.

\bibitem{Laherrere199801}J. Laherr\'{e}re, D. Sornette, Eur. Phys. J. B, \textbf{2} (1998) 528.

\bibitem{ZhangZK201001}Z.-K. Zhang, T. Zhou, Y.-C. Zhang, Physica A, \textbf{389} (2010) 179.

\end{thebibliography}
\end{document}